\documentclass[twocolumn,showpacs,preprintnumbers,amsmath,amssymb]{revtex4}

\usepackage{graphicx}

\begin{document}

\title{Counting and manipulating single electrons using a carbon nanotube transistor}

\author{An. Gruneis, M.J. Esplandiu, D. Garcia-Sanchez, and A. Bachtold$^*$}

\address{
ICN and CNM-CSIC, Campus Universitat Autonoma de Barcelona,
E-08193 Bellaterra, Spain.}

\begin{abstract}
We report on the electric measurements of an individual Au nanoparticle
with an ultra-high contact resistance of about $10^{19} \Omega$. The high-impedance measurements have
been carried out by counting the electrons that are transferred onto
the particle. In order to do this, a carbon nanotube is used as the electrode for the
supply of electrons but also as the detector of the charge transfer.
Measurements using single-electron detection allow us to
determine the separation between the electron states in the Au nanoparticle,
which is about 4~meV, consistent with the charging energy of the particle.
Single-electron detection with nanotubes offers
great promise for the study of electron transfer in highly resistive molecular systems.

\end{abstract}

\vspace{.3cm} 

\date{ \today}
\maketitle

Detection and manipulation of individual electron
charges are among the ultimate goals of nanoscale
electronics.
It holds promise for ultra-low dissipative circuits as well as for information processing
in highly-resistive molecular circuits. Carbon nanotube transistors \cite{tans,martel}
offer unique opportunities for single-electron detection. Nanotubes have ultra-small cross-sections and
their conducting electrons are located at the tube surface.
These advantages have been exploited for the sensing of
chemical gas \cite{Kong2,Collins} and
biological probes \cite{besteman,Star}. Nanotubes have also been used to detect packets of multiple electrons
transferred from the nanotube onto a particle, though the precise number
of electrons in the packet could not be measured \cite{marty}. Single-electron
detection has been resolved
for electrons hopping onto defects randomly trapped in the substrate \cite{liu,ishigami,lin,Durkop,peng}. These
single-electron processes remain however poorly controlled. Different defects can be probed
in parallel, which makes it difficult to assign different electron states to a same defect.
As a result, electron properties of such devices can only be partially characterized.

In this letter, we demonstrate for the first time single-electron detection
on a nanosystem that is not a defect \cite{liu,ishigami,lin,Durkop,peng},
namely a gold nanoparticle. This well-defined device
allows for the access of dozens of electron states of the particle, but also for
the full electron characterization of the device. In contrast with
previous works \cite{marty,liu,ishigami,lin,Durkop,peng}, we can determine
the energy separation between the electron states, which is found to be
$\simeq 4$~meV, consistent with the charging energy of the particle.
We can also determine the resistance between the particle and the nanotube,
which is about $10^{19}~\Omega$ and results in a transfer rate as low as
$\simeq0.001$~s$^{-1}$. Such a low transfer
rate, together with the well-defined device, allow us to inject and extract different
electrons from the particle
in out-of-equilibrium conditions, and to monitor the electron number
decay in time.

\begin{figure}
\includegraphics[width=7cm]{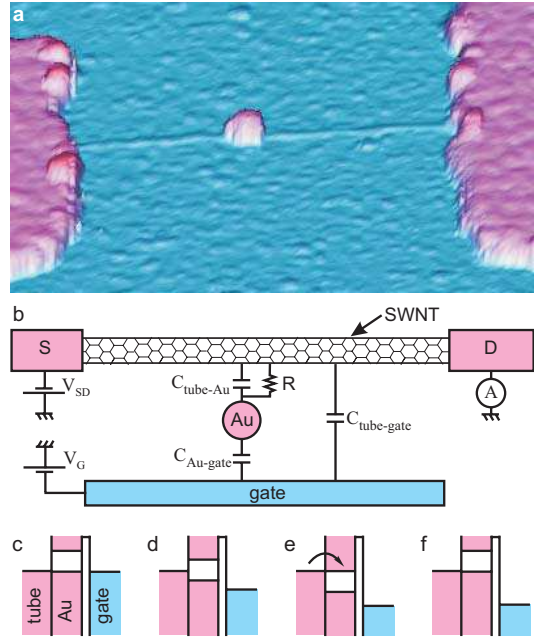}
 \caption{(color online) Device geometry.
(a) Atomic force microscopy image of a Au nanoparticle placed on top
of a SWNT, which is contacted by 2 metal electrodes. The separation between the
metal electrodes is 600~nm. (b) Schematic of the measurement
setup. The tube conductance is always measured with $eV_{SD}<kT$.
(c-f) Schematics of the potentials in the nanotube and the
particle as the gate potential is swept down. Each time
an empty energy level of the particle matches the
electrochemical potential of the tube, an electron is
transferred onto the particle, which is detected by the
nanotube transistor.
 }
\end{figure}

Carbon nanotube transistors are fabricated by means of
standard nanofabrication techniques.
Tubes are grown by chemical vapor deposition \cite{kong}
on a doped Si wafer with a 1 $\mu$m thermal silicon oxide
layer. They are electrically contacted to Cr/Au
electrodes patterned by electron-beam lithography. Gold
nanoparticles are deposited onto the wafer from a suspension in
water that consists of gold
chloride and trisodium citrate. A $\simeq30$~nm diameter particle
is positioned on top of the
tube by atomic force microscopy manipulation (Fig. 1 (a,b)).

The transfer of electrons onto the particle can be detected by measuring
the conductance $G_{tube}$ of the nanotube while sweeping
the gate voltage $V_G$ (Fig. 2(a)), as the tube conductance
is extremely sensitive to the presence of electric charges.
As $V_G$ is swept from -4 to -1~V, the conductance
is turned off as for typical p-doped semiconducting SWNTs \cite{tans,martel}.
Moreover, we have observed 35 abrupt
conductance jumps (vertical red bars) that indicate discrete electron transfers
from the nanotube into the particle. Each extra electron
in the particle changes the electrostatic potential in the
particle and, in turn, the charge density $\rho_{tube}$
in the nanotube,
which shifts the conductance $G_{tube} \propto \rho_{tube}$ horizontally in $V_G$.

\begin{figure}
\includegraphics[width=8cm]{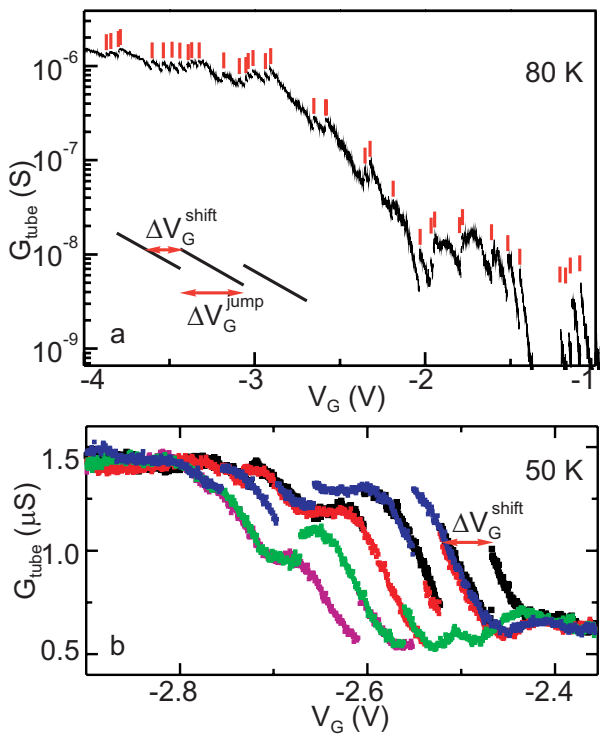}
 \caption{(color online) Detection of single electrons.
(a) Tube conductance as
the gate voltage $V_G$ is swept from -4 to -1 V. Vertical red bars indicate
conductance jumps. The recording time was 50 minutes.
(b) Tube conductance as a function of $V_G$. Each
color corresponds to a different scan. The sweep rate of $V_G$ is the same for
the different scans. The magenta curve is shown for $V_G$
between -2.9 and -2.55~V.
 }
\end{figure}

As the measurement is repeated, conductance jumps appear at
different gate voltages, see Fig. 2(b). This indicates that
electron transfers occur stochastically in time; see also
below. Remarkably, repeated measurements fall
on curves that are periodically spaced in gate voltage with a period of
about $V_G^{shift}=$60~mV. The same spacing is observed at
all the $V_G$'s from the on- to the off-conductance (from -4 to -1~V; see Fig. 2(a)). This periodicity
suggests that adjacent curves differ by one electron in the Au
 particle, and, in turn, that the observed jumps correspond
to transfers of single electrons. Measurements on a second device
yield a period of about 40~mV.

Measurements on devices without an Au particle look very different.
Most often, no conductance jumps are observed at all. For some devices, jumps can be
detected, but their number remains very low, and no period in gate voltage can be assigned.
Those jumps are attributed to uncontrolled charge traps at defects.

The mechanism that controls electron transfer onto the particle has a lot in common with what happens
in a single electron transistor \cite{sohn}. Adding an electron to the particle
costs the Coulomb charging energy $E_C=e^2/(C_{tube-Au}+C_{Au-gate})$ (represented by a gap
in Fig. 1(c)). By reducing the gate
potential $E_G \propto -eV_G$, the potential of the Au particle $E_{Au}$ goes down according to
Kirchhoff's laws (Fig. 1(d)). This is described by the first term in
\begin{equation}
E_{Au}=\frac{C_{Au-gate}}{C_{Au-gate}+C_{tube-Au}}E_G+E_CN
\end{equation}
When the tube's electrochemical potential matches the upper energy of the Coulomb
gap in the particle (Fig. 1(e)), an electron can be transferred from the tube onto the particle, and
the electron number $N$ in the particle is increased by one. This shifts $E_{Au}$ by the
amount $E_C$ (Fig. 1(f)), which blocks the transfer of the next electron.
In contrast to previous single electron transistors \cite{sohn}, the transfer rate is slow enough to prevent
the last electron from tunnelling out from the particle by continuously sweeping down the gate potential.

We will now look at the time dependence of the electron transfers. For
this purpose, the gate voltage is set at a fixed value while measuring the
tube conductance, Fig. 3. At 50~K, the tube conductance fluctuates
between two values on a time scale of several hundred
seconds. We attribute the two level fluctuations to an electron going
back and forth into the Au  particle due to thermal
excitation, and thus changing the number of electrons
between $N$ and $N+1$. As the temperature is increased to 150~K,
the tube conductance fluctuates between three levels, i.e.
between $N$, $N+1$, and $N+2$ (Fig. 3(b)). The
mechanism is schematized in the insets of Fig. 3.

\begin{figure}
\includegraphics[width=7cm]{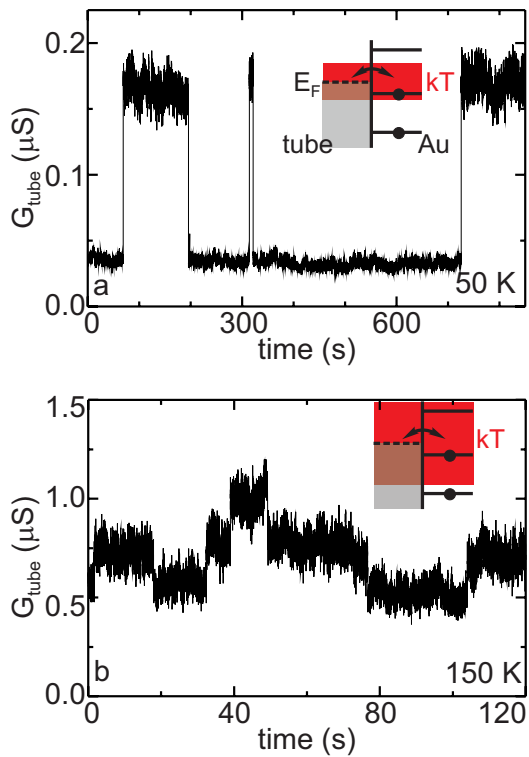}
 \caption{(color online) Fluctuations of the electron number due to thermal excitation. (a)
Tube conductance as a function of time at 50~K for $V_G=-1.35$~V.
(b) Tube conductance as a function of time at 150~K for $V_G=-1.2$~V.
The conductance experiences two levels at 50~K and three levels at 150~K. Note
that an extra level can appear at some other $V_G$'s.
We attribute the extra level at 50~K to
the electrochemical potential of the tube that matches the center of the Coulomb gap.
 The number of observed levels is on
average 2.01 at 50~K and 3.1 at 150~K. The insets show the energy levels in the
tube and in the Au particle for different numbers $N$ of electrons. The thermal
energy is shown in red.
}
\end{figure}

The fluctuations of $N$ due to thermal excitation provide
information on the energy separation $E_C$ between electron states of the Au
 particle. The two level fluctuations at 50~K
 suggest that $E_C$ is about $kT$, i.e. $\simeq4$~meV
(see inset of Fig. 3(a)).
Taking $E_C \approx 50$~K gives $C_{tube-Au}+C_{Au-gate} \approx 38$aF,
which is reasonable when considering that the self-capacitance of a
sphere $4\pi \varepsilon _r \varepsilon _0 r_{Au}= 7$ aF with
$\varepsilon _r=4$ and $r_{Au}=15$~nm has the same order of magnitude.

We will now exploit the low rate of the charge transfer
in order to manipulate the number of electrons on the particle
 in out-of-equilibrium conditions. In order to do this, the gate voltage is rapidly
swept to change the potential of the Au particle,
which positions different empty (occupied)
electron levels of the Au particle below (above) the Fermi energy
(schematic of Fig. 4(a)). The number
of empty levels depends on the amplitude of the gate voltage sweep.
Measuring the tube conductance versus time allows us to monitor the decay
towards equilibrium of the electron system (Fig. 4(a)).
As discussed previously, each conductance jump corresponds to the transfer of
one electron.

\begin{figure}
\includegraphics[width=7cm]{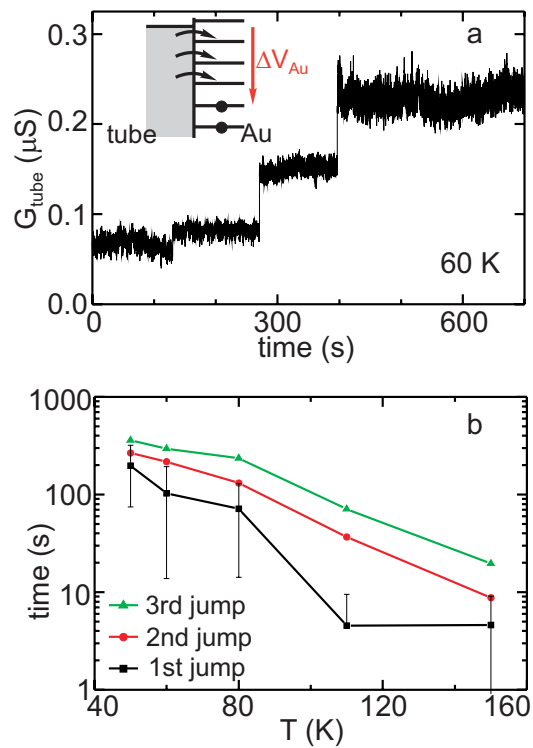}
 \caption{(color online) Manipulation of single electrons.
(a) Tube conductance as a function of time after having rapidly increased the gate
voltage by 0.4~V from -1.6 to -1.2~V. This results in the reduction of the potential in
the particle as shown in the inset. The energy levels in the particle
correspond to different numbers $N$ of electrons. Note that the conductance height
differs for the three jumps,
which is attributed to the variation of the slope of $G_{tube}(V_G)$.
(b) Average time of the first, second, and third jumps as a function
of temperature. The standard error is shown for the first jump.
 }
\end{figure}

The transfer rate changes as the temperature varies (Fig. 4(b)).
The average time for the 3 jumps in Fig. 4(a) is shown to
increase dramatically as the temperature is reduced. This suggests
that the transfer is not a simple tunnel effect, but that
transferred electrons may interact with phonons or other
electrons. In principle, such transfer rate measurements can
allow us to distinguish between thermally activated and Luttinger liquid-like
(power law dependence) behaviours of electrons in the nanotube \cite{bockrath,yao,gao}.
The origin of the temperature dependence, however, is beyond the
scope of this paper and will be left for further studies.

The measurements above allow us to estimate the electron resistance $R$ between the Au
particle and the nanotube. The average jump time $\tau$ is
\begin{equation}
\tau \approx RC_{tube-Au}
\end{equation}
This relation can be seen as the resistance given
by the voltage $e/C_{tube-Au}$
divided by the current $e/\tau$. Since $C_{tube-Au}\approx e^2/E_C$,
we get $R\approx10^{19}~\Omega$ at 50~K when $\tau \approx 200$~s. Such a
resistance is six orders of magnitude higher than what conventional electronics can cover.
The resistance may originate from a
gap of a few nm's between the
tube and the particle or from adsorbate layers at the tube-particle
interface. This resistance could not be measured in previous
single-electron detection experiments with nanotubes \cite{marty,liu,ishigami,lin,Durkop,peng}.

The device can be further characterized by
considering the electric circuit in Fig. 1(b).
This circuit has been analyzed for silicon single electron memory
using a small floating gate, which represents the ultimate miniaturization
of a flash memory \cite{Yano,Durkop}. In this model, the conductance jumps in Fig. 2(a)
are on average separated by $\Delta V_G^{jump}$ and the adjacent
curves in Fig. 2(b) by $\Delta V_G^{shift}$ with

\begin{equation}
\Delta V_G^{jump}=\frac{e}{C_{Au-gate}}
\end{equation}
\begin{eqnarray} \label{l2}
\Delta V_G^{shift} &=& e\frac{C_{tube-Au}}{C_{tube-Au}+C_{Au-gate}}
\\ \nonumber & \times& (C_{tube-gate}+\frac{C_{tube-Au}C_{Au-gate}}{C_{tube-Au}+C_{Au-gate}})
\end{eqnarray}
By taking the mean value for $\Delta V_G^{jump}=85$~mV, $\Delta V_G^{shift}=60$~mV, and
$E_C \simeq 4$~meV, we get $C_{Au-gate}=1.8$~aF, $C_{tube-Au}\approx
30$~aF, and $C_{tube-gate}\approx 1$~aF. Those values are
reasonable considering the device geometry. Indeed,
$C_{Au-gate}$ can be roughly estimated by half the capacitance between
two concentric spheres, which is $2\pi \varepsilon _r \varepsilon _0 (1/r_{Au}-1/h)^{-1}\simeq 3$~aF
with $h$ being the oxide thickness. $C_{tube-gate}$ is expected to be slightly less than
half the capacitance between
two coaxial cylinders, which is $\pi \varepsilon _r \varepsilon _0 L/\textmd{ln}(h/r_{tube})\simeq 9$~aF
with $L$ being the tube length.

The capacitance $C_{Au-gate}$ quantifies the coupling between the Au
 particle and the gate. $C_{Au-gate}$,
which is 1.8~aF, is remarkably large when considering that
the gate is 1~$\mu$m away from the Au particle.
Compared to previous experiments on Au particles directly
contacted to metal electrodes, the same
coupling can be achieved provided that the separation between the gate and the
Au particle is reduced to 2-3 nm \cite{Magnus,Bolotin}. This is because most of
the electric field in the latter case is screened by the metal
electrodes.
Overall, our device layout
enables an efficient coupling, which allows access
to a broad range of energy levels by sweeping the
gate voltage. This is especially interesting for future studies
on organic and biological molecules, since the large energy separation
between the levels has often limited
access to only one level \cite{Park,Liang}.

We will now compare our work to other existing single-electron detectors, which
are devices microfabricated in metal or semiconducting material working at millikelvin
temperatures \cite{lu,Elzerman,bylander,gustavsson,fujisawa}.
The operation temperature (up to 150~K) of nanotube detectors is much
higher. In addition, nanotubes
are suitable for electron detection on systems that are physically different from
the detector itself, such as molecules or nanoparticles.
In contrast, microfabricated single-electron detectors so far have only probed systems structured in the same
semiconducting or metal material than the detector. Moreover, these detectors are much
larger in size, which is unpractical for addressing molecules.

In conclusion, single-electron detection with a nanotube transistor has been used
to electrically probe a gold nanoparticle with an ultra-high contact resistance of
about $10^{19}~\Omega$. This is remarkable, since such a resistance is
six orders of magnitude higher than
what conventional electronics can cover. We have shown how single-electron detection with nanotubes
can be used to extract important information about the Au particle, such as
the energy separation between the electron states.
Single-electron counting with nanotubes offers great promise for future studies
on organic molecules, biomolecules, or
semiconducting particles, which most often are highly resistive. Interestingly, electron states of those systems
are expected to depend not only on the charging energy, but also
on the molecular levels.
Single-electron photoelectric effects can also be investigated, for instance,
in CdSe particles \cite{Gudiksen} as well as charge transfer in
biomolecules involved in photosynthesis
and respiration activities
\cite{Gray}. This technique may also provide information
on internal electron transfer events that occur within complex molecular systems.

We thank P. Gambardella, B. Placais, S. Sapmaz and P. Gorostiza for discussions.
The research has been supported by an EURYI grant
and the EU funded project CARDEQ (FP6-IST-021285-2).

$^{*}$ corresponding author: adrian.bachtold@cnm.es

\end{document}